# THE INVESTIGATIONS OF BEAM EXTRACTION AND COLLIMATION AT U-70 PROTON SYNCHROTRON OF IHEP BY USING SHORT SILICON CRYSTALS

A.G.Afonine, V.T. Baranov, V.M.Biryukov, V.N. Chepegin, Y.A.Chesnokov, Y.S.Fedotov, V.I.Kotov, V.A.Maisheev, V.I.Terekhov, E.F.Troyanov, Institute for High Energy Physics, Protvino, Russia; A.Drees, D.Trbojevic, BNL, Upton NY, USA; W.Scandale, CERN, Geneva, Switzerland; M.B.H. Breese, Singapore University; V. Guidi, G. Martinelli, M. Stefancich, D. Vincenzi, University of Ferrara & INFN, Ferrara

*Abstract*

The new results of using short (2-4mm) bent crystals for extraction and collimation of proton beam at IHEP 70 Gev proton synchrotron are reported. A broad range of energies from 6 to 65 GeV has been studied in the same crystal collimation set-up where earlier the extraction efficiency of 85% was obtained for 70 GeV protons using a 2-mm Si crystal. The new regime of extraction is applied now at the accelerator to deliver the beam for different experimental setups within the range of intensity 10E7-10E12ppp.

## 1 INTRODUCTION

The use of bent crystals for beam extraction in accelerators is under development at several laboratories. [1,2,3]. The advantages of this method are the ease of its realisation, feasibility of its simultaneous work with collider regime or with internal targets, and the absence of intensity pulsations because no resonant blow-up of the beam is needed to direct the beam onto the crystal for extraction. The crystal has a minimal "septum width", and is very convenient even for application in a beam-loss localisation system as a coherent scatterer [4].

Currently, bent crystals are largely used for extraction of 70-GeV protons at IHEP (Protvino). A bent crystal (5 mm Si) is installed into Yellow ring of the Relativistic Heavy Ion Collider where it channels Au ions and polarised protons of 100-250 GeV/u as a part of the collimation system [5].

A collaboration of researchers working at the 70-GeV accelerator of IHEP has recently achieved a substantial progress in the parameters of crystal-assisted beam deflection: an extraction efficiency larger than 85% has been obtained up to intensity as high as $10^{12}$ protons [4]. A major feature of such achievement was the usage of very short crystals for extraction; the crystals were selected to the minimal value foreseen by the physics of channeling [6]. Thereby, the circulating particles encountered the crystal many times and to suffered negligible divergence at each pass due to reduced scattering and nuclear interactions in the crystal. Multiple passage of particles allowed the protons to be eventually channeled by the crystal planes, leading to the experimentally recorded high efficiency.

The experimental set-up we are using for crystal extraction and crystal collimation studies has been described in some detail earlier [3,4], as well as the designs and characteristics of crystal deflectors successfully applied in crystal channeling at IHEP Protvino have been presented elsewhere [3,4]. The crystal deflector was positioned 20 m upstream of the collimator area in the U-70 ring. The crystal was carefully aligned in the direction of the incoming halo particles, which were deflected and intercepted at the entry face of the secondary collimator jaw. A hodoscope detector measured the radial beam profile observed at the entry face of the collimator jaw.

## 2 CRYSTAL COLLIMATION IN A BROAD ENERGY RANGE

In the recent experiment we have tested the same channeling crystal on the same collimation set-up in a broad energy range that was made available in the main ring of U-70 accelerator.

In our earlier paper [4] we reported a crystal collimation experiment performed at the top energy flattop, 70 GeV, and at the injection flattop, 1.3 GeV, of our machine. In the 70 GeV experiments, the extraction efficiency amounted 85% even when the entire beam stored in the 70 GeV ring was dumped onto the crystal.

This time we tested seven intermediate energies, and, importantly, it was not possible to arrange a flattop for each energy. During the acceleration part of the machine cycle, on a certain moment corresponding to the energy of the test, the beam was dumped in a short time onto the crystal. The time of beam interaction with the crystal was quite short, and amounted t=20 ms at higher energies around 65 GeV, 10 ms at the intermediate energies around 30 GeV, and only 5 ms for the lower energies down to 6 GeV as used in these tests.

Figure 1 shows an example of the radial beam profile observed at the entry face of the collimator with crystal working as primary scraper, at 45 GeV. The peak of channeled protons is well in the depth of the collimator while the multiply scattered particles are peaked at the edge. Another example is shown in Figure 2 at the energy of 12 GeV. Less channeled particles and more scattered ones are observed in the same set-up, with same crystal.

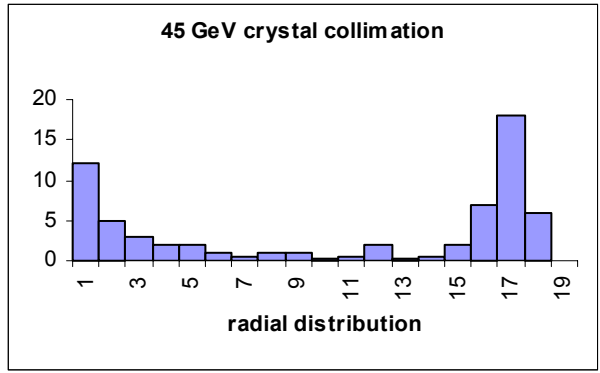

*Figure 1* The radial beam profile observed at the entry face of the collimator with crystal working as primary scraper, at 45 GeV.

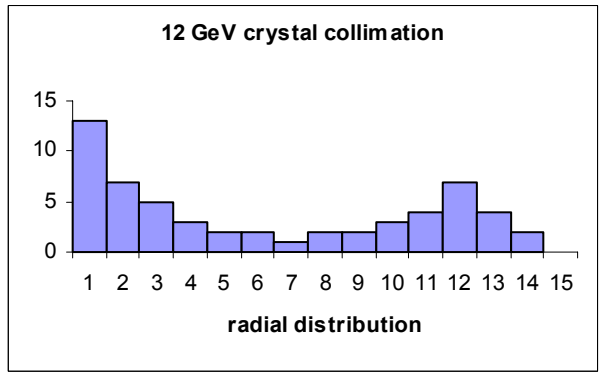

*Figure 2* The same as in Fig.1 but at 12 GeV.

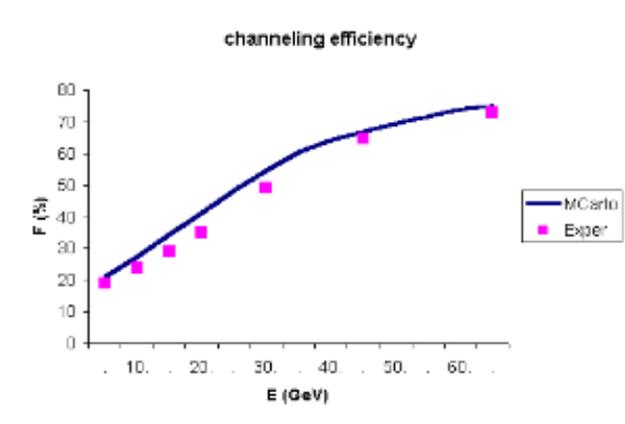

*Figure 3* The crystal collimation efficiency (channeled particles ratio to the entire beam dump) as measured and as predicted by Monte Carlo simulation.

These measurements are summarized in Figure 3 showing the ratio of the channeled particles to the entire beam dump (the *crystal collimation efficiency*) as measured and as predicted by Monte Carlo simulation.

To understand this dependence in the simulations, it was necessary to take into account the short time of beam interaction with the crystal, and especially the fact that it varied throughout the test. The most important factor for lower end of the energy range of the test was found to be the change in the primary impact factor $b_1$ of the halo protons at crystal. A rough estimate for it would be

$$b_1 \approx \frac{2\sigma_x}{t} 3\tau$$

with $\tau=5$ μs being the time of one circulation in the ring, and σ the beam size. With the energy change, the accelerated-beam size does change substantially as well.

Overall, this rough estimate suggests that $b_1$ is about 2.7 μm at 65 GeV, but as much as 36 μm at 6 GeV. The "septum width" of the crystal is likely larger than 2 μm but certainly smaller than 36μm, therefore at lower energy the contribution of the primary encounter of (undisturbed) halo proton is essential.

With that assumption on $b_1$ and the septum width of 20 μm, there is good agreement between the absolute figures of calculated channeling efficiency and the measurement.

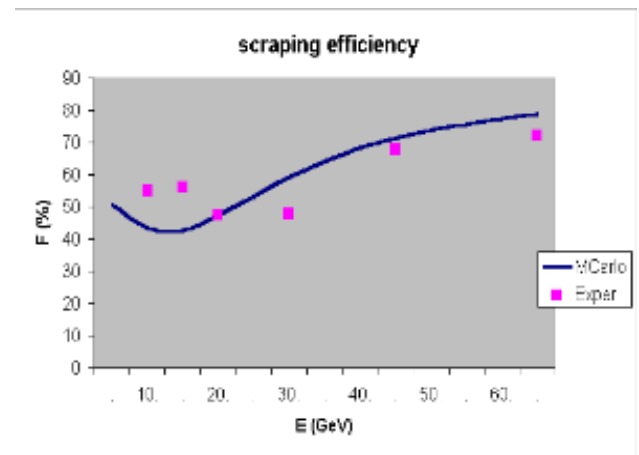

*Figure 4* The ratio of the particles intercepted by the collimator within ≥ 4 mm from the edge to all the beam loss, as measured and as predicted by Monte Carlo simulation.

Figure 4 shows the number of the particles found on the collimator entry face at ≥ 4 mm from the collimator edge, normalized to all the beam as measured and as predicted by Monte Carlo simulation. Unlike the Fig.3, which includes only the channeled peak, Fig.4 includes both the channeled peak and part of the particles scattered across the collimator face. Whereas at higher energies most of the particles are localized on the peak, at lower energies more particles are expectedly found scattered across.

## 3 CONCLUSION

Short strips of silicon crystals, bent by metallic holders with an almost constant radius, have been used to channel protons in the U-70 ring. Their channeling efficiency reached unprecedented high values in a broad range of energy, with the use of the same 2 mm long crystal deflector

These results give us a strong motivation to pursue our studies in view of proposing crystal-assisted collimation of beams in a broad energy range in order to evaluate the potential benefits for beam collimation systems in the new-generation accelerators, from spallation sources to large hadron colliders. The technique presented here is potentially applicable also in LHC for instance to improve the efficiency of the LHC cleaning system by embedding bent crystals in the primary collimators.

## ACKNOWLEDGEMENTS

This work was partially supported by INTAS-CERN Grant No. 132-2000 and RFBR Grant No. 01-02-16229.